\documentclass[aps,prl,twocolumn,amssymb,showpacs,nofootinbib]{revtex4}
\usepackage{graphicx}
\usepackage{amsmath}
\usepackage{amssymb}
\usepackage{bm}

\def\be{\begin{equation}}
\def\ee{\end{equation}}
\def\ba{\begin{eqnarray}}
\def\ea{\end{eqnarray}}

\def\l{\left}
\def\r{\right}
\def\f{\frac}

\def\sq2{\sqrt{2}}

\def\ph{\varphi}

\def\m4{m^4(\ph)}
\def\mn2{m_n^2}

\def\v5{V^{(5)}}

\begin{document}

\title{Non-Gaussian Signatures from the Postinflationary Early Universe} 
\author{Alessandra Silvestri$^{1}$}
\author{Mark Trodden$^{2}$}

\affiliation{$^{1}$Kavli Institute for Astrophysics and Space Research, MIT, Cambridge, MA 02139, USA. \\
$^{2}$Department of Physics and Astronomy, University of Pennsylvania, Philadelphia, PA 19104, USA.}
\date{\today}

\begin{abstract}
We consider contributions to non-Gaussianity of the Cosmic Microwave Background (CMB) from remnants of phase transitions in the very early universe. Such signatures can optimistically be used to discover evidence of new particle physics through cosmological observations. More conservatively they may provide an obstacle to extracting information about the non-Gaussian nature of primordial density fluctuations from any detection in the CMB. We study this explicitly by computing the bispectrum from global textures, which occur in a wide class of particle physics models.
\end{abstract}

\maketitle

Measurements of the Cosmic Microwave Background (CMB) temperature anisotropy have reached unprecedented precision, and there is now a growing interest~\cite{Komatsu:2001rj,Creminelli:2005hu,Fergusson:2006pr,Yadav:2007yy,Komatsu:2008hk} in the non-Gaussianity of the temperature field. Most of the current focus is on the exciting possibility of interpreting any detection as a signature of inflationary, or other models of the early universe. To this end considerable work has been carried out to predict the non-Gaussianity expected in the primordial 
spectrum of perturbations from inflation~\cite{Acquaviva:2002ud,Maldacena:2002vr,Bartolo:2004if,Rigopoulos:2004ba,Rigopoulos:2005ae,Lyth:2005fi,Battefeld:2006sz,Chen:2006nt,Vernizzi:2006ve,Cline:2008aw} and cyclic~\cite{Buchbinder:2007at} models. 

However, as has been discussed previously~\cite{Bartolo:2003gh,Enqvist:2004ey,Babich:2004gb,Kolb:2005ux,Brown:2005kr,Barnaby:2006km,Chen:2006xjb}, primordial effects are not the only way in which non-Gaussianity can arise in the observed anisotropy spectrum of the CMB. As photons travel from the surface of last scattering, they traverse a variety of structures, many of which may imprint secondary non-gaussianities on the spectrum. An example of this is provided by the correlation between the Sunyaev-Zel'dovich effect and weak lensing as photons pass through hot  galaxy clusters.

Another possibility is that phase transitions in the early universe can introduce a new source of non-Gaussianity, which needs to be considered when trying to differentiate primordial origins from secondary effects in any detected signal. Phase transitions in which topological defects form are particularly interesting~\cite{Gangui:2001fr}, since these may source non-Gaussianity actively throughout cosmic history, allowing for an integrated effect. Thus, even though topological defects cannot be the primary agent in structure formation, they may contribute a significant non-Gaussianity in the CMB temperature field. 

In this letter we focus on the example of a rather special class of topological defects - {\it textures}~\cite{Turok:1990gw,Pen:1993nx,Borrill:1994uh,Magueijo:1995uy}. These occur whenever a particle physics theory possesses a global symmetry group $G$ that is broken to a subgroup $H$ in such a manner that the third homotopy group $\pi_3$ of the vacuum manifold ${\cal M}\equiv G/H$ is nontrivial. A simple example is given by the Lagrangian density
\be
{\cal L}=\frac{1}{2}(\partial_{\mu}\Phi_i)\partial^{\mu}\Phi_i -V(\Phi_i) \ ,
\ee
where $\Phi_i$ ($i=1,\ldots,4$) are real scalar fields, and the choice of the symmetry breaking potential
\be
V(\Phi_i)=\frac{\lambda}{4}\left(\Phi_i\Phi_i -\eta^2\right)^2 \ ,
\ee
with $\lambda$, $\eta$ constants, breaks the symmetry group $G=SO(4)$ down to $H=SO(3)$, with $\pi_3(G/H)\cong Z$.

Textures consist entirely of gradient energy, which evolves before ultimately unwinding to topological triviality in a small region of spacetime, and radiating its energy out to infinity. Our goal here is to estimate the non-Gaussianity from textures by computing the bispectrum, illustrating the care required when interpreting any detected non-Gaussianity as a primordial signal.

As usual, the fluctuations of the temperature of the CMB can be decomposed into spherical harmonics via
\begin{equation*}
\label{deltaTspectrum}
\frac{\Delta T}{T}({\hat {\bf n}}) =\sum_{lm} a_{lm}Y_{lm}({\hat {\bf n}})\ ; \ \ \ \ 
a_{lm}\equiv\int d^2{\hat {\bf n}}\frac{\Delta T}{T}({\hat {\bf n}})Y_{lm}^*({\hat {\bf n}}) \ .
\end{equation*}
The bispectrum is then the 3-point correlation function
\be\label{bispectrum_def}
B^{m_1m_2m_3}_{l_1l_2l_3}\equiv \langle a_{l_1m_1}a_{l_2m_2}a_{l_3m_3}\rangle
\ee
and one observes the angle-averaged bispectrum 
\be\label{ang-av_bispectrum}
B_{l_1l_2l_3}\equiv\sum_{m_1,m_2,m_3}\left(\begin{array}{ccc}l_1&l_2&l_3\\m_1&m_2&m_3\end{array}\right)B^{m_1m_2m_3}_{l_1l_2l_3} \ .
\ee

Here, in order to study the non-Gaussianity that textures source, we compute the bispectrum~(\ref{ang-av_bispectrum}), elaborating on the analytical results of~\cite{Magueijo:1995uy,Gangui:1996cg}. 

Any random distribution of scalar field orientations will contain a significant amount of gradient energy, which will relax to vacuum. Textures themselves are entirely gradient energy and, due to topology, a configuration will collapse until there is sufficient gradient energy in a small enough volume that it becomes energetically favorable for the field to unwind. Both cases have an evolving gravitational potential as the configuration reaches triviality by radiating goldstone modes. As articulated in~\cite{Turok:1990gw}, the effect of a particular texture on CMB photons depends on whether the photons traverse the texture while  it is unwinding or afterwards. Thus, textures create hot and cold spots in the CMB. Simulations~\cite{Borrill:1994uh}, show an analogous effect even if the field configuration in question does not carry topological charge.

The defect network, and consequently the network of hot and cold spots, is expected to scale, i.e. to look statistically the same at any time, with characteristic length scale normalized to the horizon size. Following~\cite{Magueijo:1995uy}, we model the spot network with a distribution function
\be\label{spot_density}
 N(y)=2\,n\ln{(2)}\,(2^{y/3}-1)^2 \ ,
 \ee
 \noindent where $n$ is the average number of texture-producing configurations per horizon volume, $y(t)\equiv \log_2(t_0/t)$ and $t_0$ is the cosmic time today. (Note that we have assumed a matter dominated universe, since the most important effects come form later textures, during this epoch.) Each spot will then contribute a temperature fluctuation
 \be\label{temperature_spot}
\left.\f{\Delta T}{T}\right|_i=a_i S_i(\theta_i,y) \ ,
 \ee
where $a_i$ and $S_i$ are the brightness and the profile, respectively, of the $i$-th spot, and $\theta_i$ is the angle in the sky as measured with respect to the center of the spot. A spot appearing at time $y$ has a typical angular size $\theta_s(y)$ which is a fraction $d_s$ of the horizon angular size
 \be\label{spot_size}
 \theta_s(y)=\arcsin{\l[{\rm min}\l(1,\f{d_s}{2(2^{y/3}-1)}\r)\r]} \ .
 \ee

The contribution of the i-{\it th} texture spot to $a_{lm}$ is then
\be\label{texture_alm}
a^i_{lm} =a^i S^i_l(y){Y^*_{lm}}(\Omega_i) \ ,
\ee
where $S_{l}(y)$ is the multipole of the spot profile
 \be\label{profile_multipole}
 S^i_l(y)\equiv 2\pi\int_{-1}^1S^i(\theta_i,y)P_l(\cos{\theta_i})\,d(\cos{\theta_i}) \ .
 \ee
 
Summing over the distribution of spots in the sky from last scattering until today (and assuming no correlations between spots
laid down at different times, between spot brightness and profile, and that spot brightness doesn't depend on the time at which the spot was laid down) one finds that the global texture bispectrum takes the form
\be\label{texture_bispectrum}
B_{l_1l_2l_3}=\gamma_{l_1l_2l_3}\langle a^3\rangle{\cal I}^3_{l_1l_2l_3} \ ,
\ee
where ${\cal I}^3_{l_1l_2l_3}$ is the integral over the profile multipoles
 \be
 {\cal I}^3_{l_1l_2l_3}\equiv \int_0^{y_{\rm {ls}}}\, dy N(y)S_{l_1}(y)S_{l_2}(y)S_{l_3}(y)\,,
  \ee
with $y_{\rm {ls}}$ the value of $y$ at last scattering and where
\be\label{geom}
\gamma_{l_1l_2l_3}\equiv\sqrt{\f{(2l_1+1)(2l_2+1)(2l_3+1)}{4\pi}}\left(\begin{array}{ccc}l_1&l_2&l_3\\0&0&0\end{array}\right) \ .
\ee

Here $\langle \cdots \rangle$ is the average over simulated spot distributions. As we do not have at our disposal the full spot brightness distribution from numerical simulations, we assume that any hot(cold) spot has the same $a_{\rm hot}>0$($a_{\rm cold}<0$) corresponding to the averaged maximum(minimum) brightness of simulations. Then we may write $\langle a^3\rangle=x\langle a^2\rangle\,\langle |a|\rangle$, and introduce the parameter $x\equiv\langle a\rangle/\langle |a|\rangle$ (assumed to be $\eta$-independent) encoding the asymmetry between cold and hot spots~\cite{Gangui:1996cg}.

For simplicity we specialize to the gaussian profile,
 \be\label{gaussian_profile}
 S_l(y)=2\pi\l[1-\cos{(\theta_s(y))}\r]\rm{exp}\l[-\f{\l(l+\f{1}{2}\r)^2\theta_s(y)^2}{2}\r] \ ,
 \ee 
and focus on calculating  the contribution of textures to the equilateral bispectrum (for which $l_1=l_2=l_3\equiv l$, not to be confused with the ``equilateral shape bispectrum") on large scales.  We choose
$n=1$, $d_s=0.1$ (as suggested by simulations~\cite{Borrill:1994uh}, ensuring that we include the effects of all scalar field gradients) and $x\simeq 0.07$~\cite{Gangui:1996cg}. We leave $\langle a^2\rangle\propto \eta^4$ as a free parameter related to the symmetry breaking scale $\eta$. Note that the observed CMB angular power spectrum bounds $\eta \leq 10^{-2}M_P$ when $d_s=0.1$, (using the $C_{10}$ normalization from WMAP5~\cite{Komatsu:2008hk}). 
\begin{figure}[t]
\includegraphics[width=60mm]{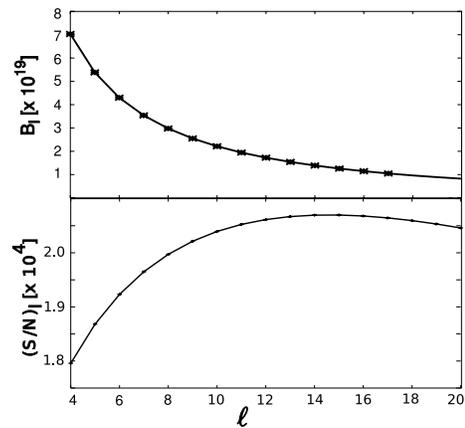}
\caption{Top panel: equilateral bispectrum for textures, $B_{l}^{\rm{text}}$ (with $\eta=2\cdot10^{-3}M_P$), as a function of $l$. Bottom panel: $S/N$ for the same as a function of $l$. Fig. \ref{StN_fig} shows that higher $\eta$ yields a higher $S/N$, with $S/N\sim O(1)$ for $\eta\sim 5\cdot10^{-3}M_P$.}\label{eq_bispectra}
\end{figure}

\begin{figure}[t]
\includegraphics[width=60mm]{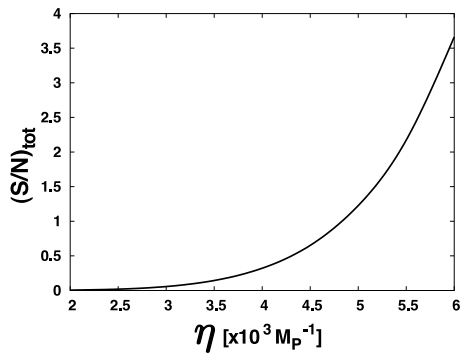}
\caption{The total S/N for the equilateral bispectrum $B^{\rm{text}}_{lll}$ of textures as a function of the symmetry breaking scale $\eta$.}\label{StN_fig}
\end{figure}

\begin{figure}[h]
\includegraphics[width=60mm]{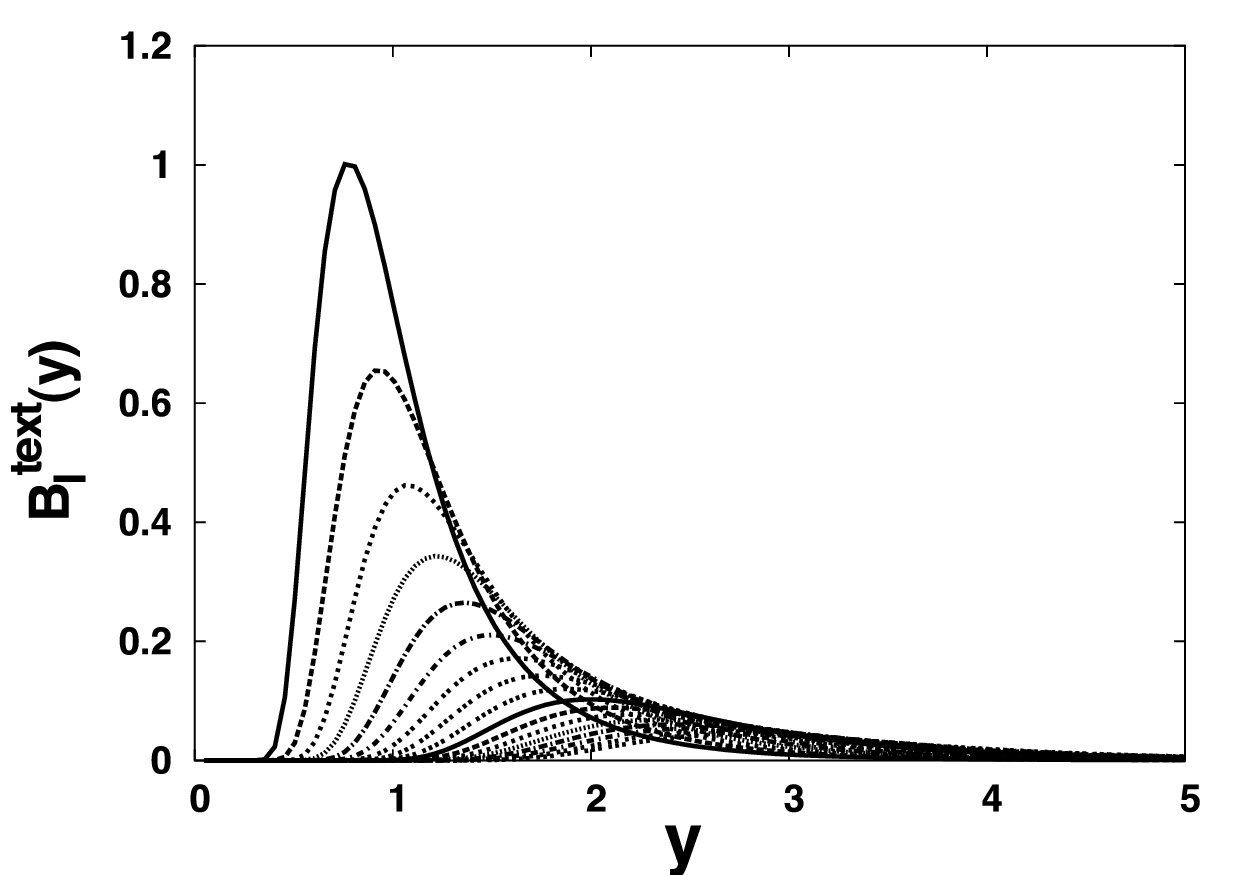}
\caption{The texture equilateral bispectrum density, $B^{\rm {text}}_l(y)$, for $d_s=0.1$, $\eta=2\cdot10^{-3}M_P$, and from top to bottom $4<l<16$. The curves have been normalized to $b_4(y_{max})$}
\label{eq_bisp_dens}
\end{figure}

We focus on the angle-averaged bispectrum (normalized as in~\cite{Gangui:2001fr}) in the equilateral case, $l_1=l_2=l_3$
\ba\label{norm}
B_l&\equiv& l(2l+1)^{3/2}\left(\begin{array}{ccc}l&l&l\\0&0&0\end{array}\right)B_{lll}\nonumber\\
&=&\sqrt{4\pi}\,\gamma_{lll}\,l\,\,B_{lll} \ .
\ea
In fig.~\ref{eq_bispectra} we plot the texture equilateral bispectrum $B^{\rm{text}}_l$, 
 finding it to be a monotonically decreasing function of multipole index; i.e. textures contribute more non-Gaussianity on larger scales (in agreement with~\cite{Magueijo:1995uy}, in which the angular spectrum ${\it C}_l$ was studied.). In the same figure we plot the theoretical signal-to-noise ratio $S/N$ for the equilateral bispectrum $B_{lll}$, which is also a monotonically decreasing function of $l$. For the variance of $B_{lll}$, we use the equilateral version of an expression in~\cite{Gangui:2000gf}, valid for mildly non-Gaussian distributions,
 \be\label{variance}
 \sigma^2_l=\langle B^2_{lll}\rangle-\langle B_{lll}\rangle^2\approx 6\,C_l^3 \ ,
 \ee
where $C_l$ includes the power spectrum of the detector noise, calculated using the analytical expression of~\cite{Knox:1995dq}, and we have included the contribution from textures. Therefore we have $\l(S/N\r)_l\approx|B_{lll}/C_l^{3/2}|$.
In fig.~\ref{StN_fig} we plot the total $S/N$ as a function of $\eta$. As the symmetry breaking scale increases, the signal increases at a bigger rate than the associated noise, therefore the ratio $S/N$ is found to be an increasing function of $\eta$, with $S/N\sim O(1)$ for $\eta\sim 5\cdot10^{-3}M_P$. 
To gain some physical insight, we may look at the {\it angle-averaged bispectrum  density} for the equilateral case:
\be\label{reduced_density}
B^{\rm{text}}_l(y)\equiv \sqrt{4\pi}\,\gamma^2_{lll}\,l\, \langle a^3\rangle\,N(y)\,S_l(y)^3 \,.
\ee
It is worth noticing that the geometrical factor~(\ref{geom}) enters eq.~(\ref{reduced_density}) quadratically, therefore there are no oscillations associated with the alternating sign of the 3j-wigner symbol.
In the case of gaussian profiles this reads
\ba\label{reduced_density_gaussian}
B^{\rm{text}}_l(y)&\equiv&32\pi^{5/2}\,\gamma^2_{lll}\,l\,\langle a^2\rangle^{3/2}x\,n\,\ln(2)\,\l(2^{y/3}-1\r)^2\nonumber\\
&&\cdot\l[1-\cos(\theta_s(y))\r]^3e^{-\f{3}{2}\l(l+\f{1}{2}\r)^2\theta_s^2(y)} \ ,
\ea
and in fig.~\ref{eq_bisp_dens} we plot $B^{text}_l(y)$ as a function of the ``time" variable $y$. The width of the curves increases as a function of the multipole index $l$, while the height of the peaks decreases as a function of $l$. The tilt of the bispectrum in fig.~\ref{eq_bispectra} depends on the balance of these two effects. It is worth pointing out that, as $l$ decreases, not only do the curves narrow, but the peaks shift to smaller values of the time $y$; i.e. lower multipoles receive most contribution from the most recently formed textures (A similar effect was found in~\cite{Magueijo:1995uy} for the power spectrum itself.)

\begin{figure}[t]
\includegraphics[width=60mm]{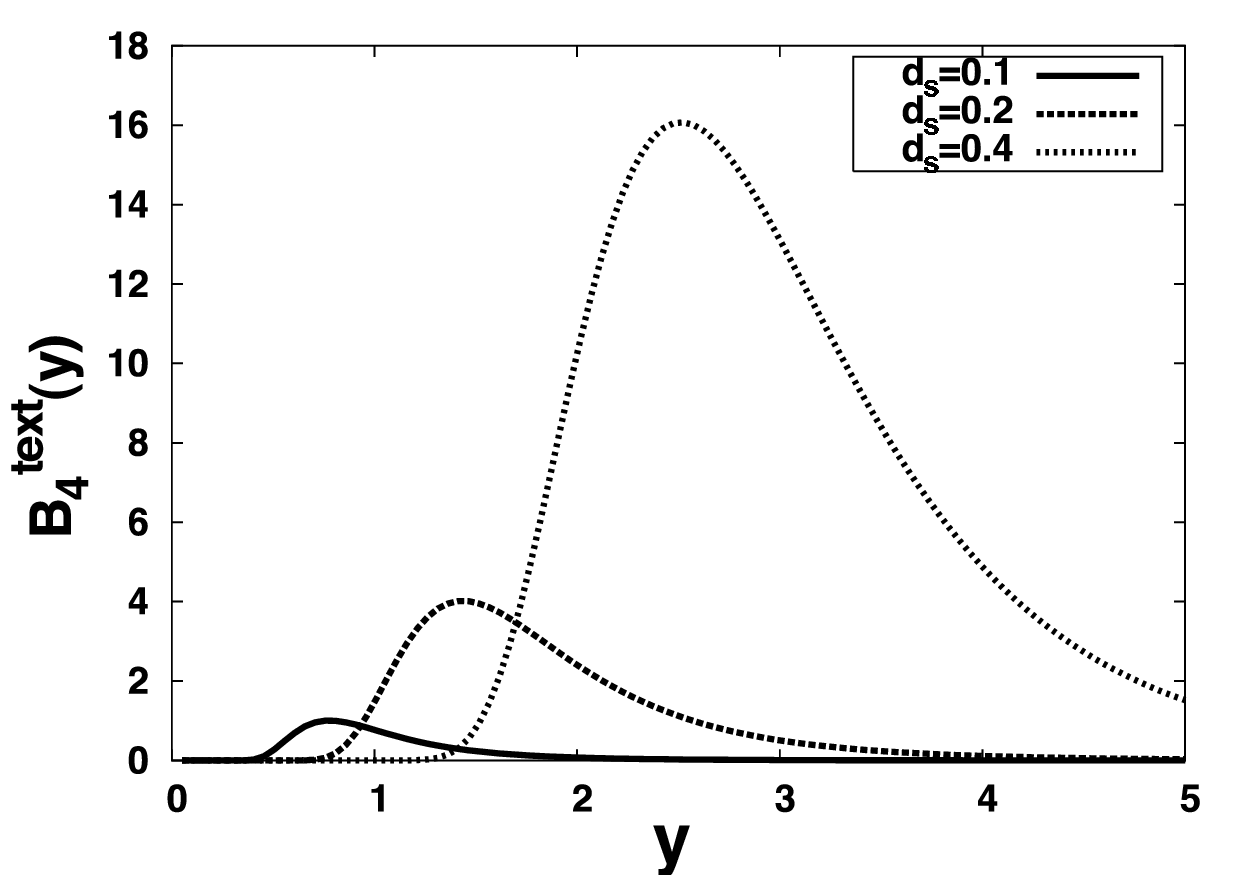}
\caption{Equilateral bispectrum density for $l=4$, $B^{\rm {text}}_4(y)$, for three different values of the density parameter; $d_s=0.1$ (solid line), $d_s=0.2$ (dashed line) and $d_s=0.4$ (dotted line) (normalized to the case $d_s=0.1$). In all cases $\eta=2\cdot10^{-3}M_P$.}
\label{b_2_ds}
\end{figure}

It is interesting to study $B_l^{\rm text}$ as a function of the impact parameter $d_s$ at fixed multipole $l$. We do this in fig.~\ref{b_2_ds} for $l=4$ for three different values of the impact parameter. As can be seen, the position of the peak shifts to higher values of $y$ for increasing values of $d_s$. Thus, the larger the value of $d_s$, the earlier the time at which the main contribution to the non-Gaussianity is generated. It is worth noting that, as $d_s$ is changed, the position of the peak changes such as to keep the angular size of the spot nearly fixed (for fixed $l$). This is as one might expect, since the spots are uncorrelated. Thus, the main contribution to non-Gaussianity at a given multipole $l$ arises at times at which the textures have an angular size of order the angular size of the multipole $\simeq 4\pi/l(l+1)$. 

\begin{figure}[t]
\includegraphics[width=60mm]{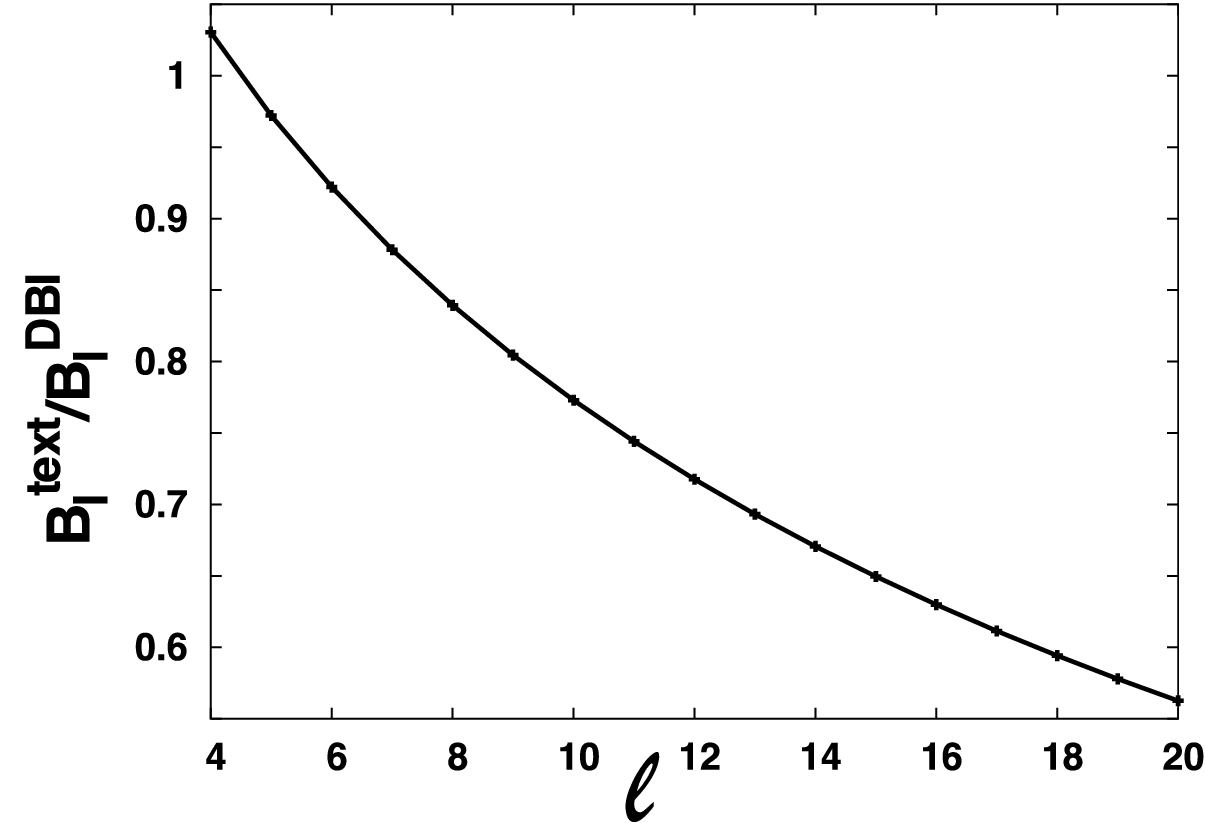}
\caption{Ratio of the bispectrum from textures, for $\eta=7\cdot10^{-3}M_P$, to the bispectrum from DBI inflation~\cite{Alishahiha:2004eh}, with speed parameter $\gamma=30$, as a function of the multipole index.}
\label{text2DBI}
\end{figure}

Post-inflationary early universe physics may generate non-Gaussian signatures in the CMB. Such signatures may be useful in identifying new physics in upcoming missions, but it is also important to understand them in order to be able to interpret any signal as evidence for primordial physics. 
As a concrete example we have computed the cosmic texture bispectrum, in a way differing from previous analyses~\cite{Jaffe:1993tt,Magueijo:1995uy,Phillips:1995nu,Gangui:1996cg,Gangui:1997hx,Phillips:2000af,Cruz:2007pe,Cruz:2008sb,Cruz:2004ce} in the analytical modeling and the use of updated datasets and current bounds on the allowed texture energy scale. Our result may be compared to the primordial bispectrum from, say, DBI inflation, which is of the equilateral form~\cite{Alishahiha:2004eh}. In fig.~\ref{text2DBI} we show an example of the ratio of these bispectra, which is a nontrivial function of the multipole index $l$, since the individual bispectra have different shapes.
Observational constraints~\cite{Komatsu:2008hk} are commonly quoted in terms of the quantity $f_{\rm NL}$. We may estimate the equilateral $f_{\rm NL}$ from textures by comparing $B_l^{\rm text}$ to that obtained from the $3$-point correlation function of the curvature perturbation
\be\label{3point}
\langle \xi_{k_1}\xi_{k_2}\xi_{k_3}\rangle=(2\pi)^7\delta^{(3)}\left(\sum\vec{k}_i\right)\left(-\frac{3}{5}f_{\rm NL}\Delta_{\xi}^2\right)\frac{4\sum_ik_i^3}{\Pi_i2k_i^3}
\ee
evaluated on an equilateral triangle $k_1=k_2=k_3$. The resulting estimate is a function of the symmetry breaking scale, with $|f_{\rm NL}^{\rm eq,text}|\simeq 5\cdot 10^{14}(\eta/M_P)^6$ with a value of  O(10) for $\eta\sim 5\cdot 10^{-3}M_P$.

\acknowledgments
We thank Rachel Bean, Levon Pogosian and Alex Vilenkin for discussions and an anonymous referee for useful suggestions. This work was supported by NSF grants PHY-0653563 (MT) and AST-0708501 (AS).

\end{document}